\documentstyle[12pt,aasms,flushrt]{article}

\def\be{\begin{equation}}
\def\ee{\end{equation}}

\begin{document}

\title{\bf The role of chaotic resonances in the solar system}

\medskip

\author{N. Murray$^1$ and M. Holman$^2$}

\medskip

\affil{$^1$Canadian Institute for Theoretical Astrophysics, \\
60 St. George St.,  University of Toronto, Toronto, Ontario M5S 3H8, Canada}

\medskip

\affil{$^2$Harvard-Smithsonian Center for Astrophysics,\\
60 Garden St., Cambridge, MA 02138, U.S.A.}

\medskip

\vfil\eject
{\bf Our understanding of the solar system has been revolutionized by
the finding that multiple planet systems are subject to chaotic
dynamical processes. The giant planets in our solar systems are
chaotic, as are the inner planets (independently).  Recently a number
of extrasolar planetary systems have been discovered; some of these
systems contain multiple planets.  It is likely the some of the latter
are chaotic.  In extreme cases chaos can disrupt some orbital
configurations, resulting in the loss of a planet. The spin axes of
planets may also evolve chaotically. This may adversely affect the
climate of otherwise biologically interesting planets.  }

\vfil\eject

\noindent

Observers have targeted solar-type stars in their searches for
planets, partly motivated by the belief that these stars might harbor
Earth-like planets.  Although present detection
techniques are unable to detect terrestrial-mass planets, they are
sensitive to Jupiter-mass bodies. The original notion was that because
all of these solar systems should share a common planet formation
process they would result in similar collections of planets; large
planets might be accompanied by smaller counterparts.  However, the
wide variety of orbits of the planets detected to date shows that
our knowledge of planet formation is incomplete.  Nevertheless, while
the formation processes in other solar systems might be quite different
from those that shaped our own, the long-term dynamical evolution is
governed by the same principles.

These principles have been known since Newton, but their consequences
for our solar system still surprise researchers. For nearly 300 years
astronomers believed that the orbits of the planets were regular and
predictable.  After an initial phase of formation, the number of
planets, asteroids, and comets was fixed.  Planets neither escaped the
solar system nor collided with each other.  The discovery of chaos
destroyed this traditional picture.

Our solar system provides a plethora of examples of chaotic
motion. The theory of chaos has been used to explain, and in some
cases predict, the location and extent of gaps in the asteroid and
Kuiper belts\cite{giffen}$^-$\cite{mdl}.  The theory
predicts that irregularly-shaped satellites such as Hyperion tumble
chaotically\cite{wpm}.  The obliquity of Mars undergoes large
excursions\cite{wr}; these are in fact chaotic\cite{tw}$^,$\cite{lr}.  Even the
obliquity of the distant future Earth may undergo chaotic
evolution in 1.5-4.5~Gyr\cite{ward82}$^-$\cite{surgy_laskar}.
Furthermore, we have recently shown that the chaos in the orbits of
the giant planets, first observed by Sussman and Wisdom\cite{sw92} is due to
a three-body resonance among Jupiter, Saturn, and Uranus\cite{mh99};
the locations of these planets cannot be predicted on a time scale
longer than a few tens of millions of years.  This provides analytic
support to earlier numerical experiments that demonstrated that our
planetary system evolves chaotically\cite{sw92}$^,$\cite{laskar89}.  The
worried reader may find some comfort; the accompanying analytic theory
predicts that no planet will be ejected before the sun dies.

{\large {\bf The rigid pendulum}}

Despite the variety and complexity of the applications mentioned
above, we can introduce many of the concepts in solar system dynamics
using the pendulum: phase space structure, periodic motion, and
stability.  To describe the state of the pendulum we must specify both
its position (such as the angle from the downward vertical, $\theta$,
see figure~\ref{pendulum}) and velocity 
(such as the time rate of change of the angle,
$\dot\theta$, or the momentum $p=ml\dot\theta$, where $m$ is the effective mass
of the pendulum and $l$ is the length).  The structure of the
phase space is built around {\it fixed points} (see
figure~\ref{pendulum_phase}). When the pendulum is hanging straight
down and motionless we have a fixed point.  There is another fixed
point when the pendulum is straight up and motionless.  Readers
equipped with a pencil can immediately see that these two points are
different: dangling a pencil loosely between two fingers is easy, but
balancing a pencil upright challenges the most dextrous.

If a pendulum at the downward fixed point is nudged it begins to
oscillate or {\it librate} back and forth; this fixed
point is said to be stable.  If a pendulum momentarily balanced on
its head is gently nudged it swings down and back up on the other side;
this fixed point is unstable. If the pendulum is pushed forcefully
enough it will rotate periodically. Thus, we see three main regions of
the phase space: circulation in one direction, libration, and
circulation in the other direction.  These three regions are separated
by the curve that connects the unstable fixed point to itself.  This
curve forms the ``cat's eye'' in figure~\ref{pendulum_phase}, and is called the
separatrix.  All trajectories inside the separatrix are said to be in
{\it resonance}.  Trajectories well outside the ``cat's eye'' have
enough kinetic energy that they are unaffected by the gravitational potential.
They rotate with an almost fixed angular frequency.  In contrast
trajectories inside the resonance are dominated by the potential; they
librate.  The average value of $\dot\theta$ in the  resonance is
zero, which is a generic signature of resonant phenomena.

The separatrix is the cradle of chaos. Suppose we reach in a hand and just
nudge a moving pendulum in an attempt to change the character of its
motion.  If the pendulum is far from the separatrix (either librating
near the stable fixed point or rotating) the nudge will not change the
character of the motion. However, close to the separatrix the pendulum
nearly stands motionless on its head once every libration or rotation
period. A nudge can then change rotation into libration, libration
into rotation, or rotation in one direction into rotation in the other
direction. Aside from the separatrix, all the orbits in
figure~\ref{pendulum_phase} lie on closed curves.  These curves are
called invariant curves; similar curves appear in most dynamical systems. 

In order for chaos to occur there must be at least two interacting
oscillators. In the solar system that interference is supplied by a
third body, the invisible hand made immanent.  This interference
manifests itself as zones of chaotic motion, which are the ghosts of
the departed separatrices.  That phase space would harbor regions
where trajectories are largely quasiperiodic and regions where
trajectories are largely chaotic is a feature of generic dynamical
systems.  This characteristic is known as a {\it divided phase
space}\cite{Chirikov60}$^,$\cite{hh}.

{\large {\bf Kirkwood gaps}}

A series of remarkable features in the asteroid belt vividly
illustrates the importance of dynamical chaos in the solar system.
The distribution of semimajor axes of the orbits of the asteroids
contains a number of distinct gaps.  These are call Kirkwood gaps, in
honor of Daniel Kirkwood\cite{kirkwood}.  It was he who first identified them and
noted that they occur at locations where the orbital period, $T$,
which depends on semimajor axis, would be of the form $(p/q) T_J$, where $T_J$ is the orbital
period of Jupiter and $p$ and $q$ are integers.  In terms of the
orbital {\it frequencies}, $n=2\pi/T$, $q n_J - p n \approx 0$; in
other words this is a resonance.  Nearly one hundred years
passed it was explained how a resonance produced a shortage of
asteroids, or a gap. The reason for this delay can be appreciated by
making a simple estimate of the strength of a resonant
perturbation. The force is proportional to the mass of Jupiter, which
is $\mu_J\sim0.001$ in units of the solar mass. Worse still, the
resonant force is zero if the orbits are circular (for a planar model). If the orbits are
non-circular, with an eccentricity $e$ for the asteroid and $e_J$ for
Jupiter, the force is proportional to $e^{|p-q|-s}e_J^s$, where $s$ is
an integer between zero and $|p-q|$. For example, for a 3:1 resonance
$p=3$ and $q=1$, three force terms are possible: $e^2$, $ee_J$, and
$e^2_J$. Since $e\sim e_J\sim 0.05$, the force exerted on the asteroid
by the resonance is smaller by a factor of $5\times10^{-5}$ than the
force exerted by the sun, in the best possible case of a first order
resonance ($|p-q|=1$).

Laplace noted even before the discovery of the Kirkwood gaps that a
resonant force adds up over many orbits, but only up to half the
libration period (the time to go half way around the cat's 
eye in figure \ref{pendulum_phase}) which is proportional to the
inverse of the square root of the resonant force. The integrated force
is then larger, roughly like the square root of $\mu_Je^{|p-q|}$. This
is still a small effect, so it is difficult to see how the asteroid
can be ejected from the resonance.

Even if the force added up over longer times, theorems on the behavior
of dynamical systems from the 1950's and '60's suggested that gap
formation might be difficult. For example, the motion of a typical
asteroid in an imaginary solar system with very small planetary masses
lies on invariant curves similar to those in
figure~\ref{pendulum_phase} (see Box 1). This type of result is known
as a KAM theorem\cite{kolmogorov}$^-$\cite{moser}. In
exceptional cases, near the separatrix of 
a resonance, the theorem does not apply.

The fact that the theorems are restricted to very small masses and
the demonstration of a divided phase space suggest
a way to evade the dilemma; perhaps for realistic values, invariant
curves at substantial distances from the separatrix are
destroyed. Numerical integrations by Giffen of a simplified model,
incorporating only a single planet (Jupiter) on a fixed ellipse,
showed that the invariant curves were destroyed in the vicinity of the
2:1 resonance\cite{giffen}. This removed the barrier of the KAM
theorem. Unfortunately, the $\sim$100,000~yr integrations by Giffen and
others, while establishing the presence of chaotic orbits,
did not give any hint as to how the irregular motion led to the
removal of an asteroid from the resonance\cite{giffen}$^,$\cite{fs}.

The critical advance came when Wisdom showed by direct numerical
integration that the eccentricity of small bodies placed in the 3:1
gap alternates chaotically between periods of low and moderate
eccentricity as the result of perturbations from
Jupiter\cite{wisdom82} (see figure~\ref{Wisdom31}).  This result
relied upon upon the development of an algebraic mapping that could
efficiently follow the long-term motion of asteroids in the 3:1 mean
motion resonance with Jupiter for a million years or longer.  This is
a recurring theme in solar system dynamics: more efficient algorithms
and faster computers permit longer, more accurate integrations often
reveal unexpected dynamical results.

What is the source of the chaos in the 2:1 and 3:1 mean motion
resonances?  It has long been known that each mean motion resonance is
composed of several individual resonant terms\cite{meanmotion}.  For
example in the 3:1 there are three such 
terms in the planar case, as mentioned above.  While each of these
resonances are weak, they are also close
together\cite{wisdom82}$^-$\cite{wisdom85a}$^,$\cite{goldreich}.
It is the close proximity of multiple, albeit weak, resonances which
invalidates the simple estimate for the integrated force given
above. A single resonance produces a small force integrated over a
libration period, and a simple ``cats eye'' in phase space with a
smooth separatrix. Adding a second weak resonance breaks up this
smooth separatrix in a chaotic region (see Box 1). The separatrix is
broken because the second resonance nudges the pendulum represented by the first resonance,
altering the motion, particularly of orbits that pass near the
unstable fixed point\cite{LL}. As a result the motion from one
passage near the fixed point to the next is uncorrelated.  The
resonant force is still small compared to the force exerted by the
sun, and produces only small changes over a libration period. But over
times longer than a libration period the asteroid experiences
uncorrelated forces.

As a result of these uncorrelated forces the motion in the chaotic
zone differs from that in a regular region; the chaotic motion is very
sensitive to initial conditions.  Suppose we take two copies of our
interacting oscillators, released from slightly different positions.
In one case we start near a stable fixed point, well away from the
separatrix.  Here the motion is regular and the two copies of the
apparatus separate from each other linearly (or polynomially) in time
(see figure~\ref{separation}).  In the other case we start the two
copies near an unstable fixed point.  Now the separation increases
exponentially in time.  The time scale for this separation is called
the Lyapunov time, $T_L$.

We can estimate the Lyapunov time of an asteroid placed artificially
in the 3:1 mean motion resonance. There are only two relevant time
scales in the problem: the libration period and the precession period.
From our argument above, large-scale chaos occurs when the separation
of the islands is comparable to their widths\cite{Chirikov}, that is,
when the resonances overlap. Another way to say this is that large
scale chaos occurs when the libration and precession periods of a
resonant asteroid are similar.  In that situation the Lyapunov time
must be equal to the libration period---the only time scale in the
problem\cite{MH97}.  In the case of the 3:1 resonance the precession
period is $3\times10^4$~years while the numerically-determined
Lyapunov time is $1.4\times10^4$~years.

The two types of trajectories differ in more than their rates of
separation.  The chaotic trajectory can explore a larger volume of
phase space than is accessible by a regular trajectory.  As noted by
Wisdom, for asteroids this extra measure of freedom can be dangerous,
opening routes by which they can be ejected or cannibalized by
planets\cite{wisdom82}$^-$\cite{wisdom85b} (it's
world-eat-world out there).

So far we have discussed regular and chaotic motion as if the two
occurred in different systems, but in fact most dynamical systems
exhibit a rich mixture of regular and chaotic motion. Box 1 describes
this in more detail.

In higher order resonances, where the phase space structure is relatively simple, the
uncorrelated or chaotic forces lead to simple random walks involving
only small exchanges of energy between the planet and the asteroid,
since the semimajor axes of both bodies are nearly fixed by the
resonance condition. Not so for the angular momentum; the torques
associated with the uncorrelated forces produce a random walk in the
angular momentum of both bodies. This leads to a classic gambler's
ruin problem with the asteroid playing the part of the gambler and the
planet taking the role of the house. If either body loses most of its
angular momentum, that body's orbit becomes highly eccentric and
subject to collisions with other objects in the solar system or
ejection---the analog of bankruptcy. The planet, with its much larger
mass, has more capital (angular momentum) than the asteroid. Even
worse, the asteroid is like a gambler forced to limit his winnings to
what he can carry in his pockets; because of its small mass the
asteroid cannot absorb enough angular momentum to produce a
substantial change in the orbit of the planet. The two trade angular
momentum back and forth, but when the inevitable happens and the
asteroid loses its small stock of angular momentum, it is abruptly
escorted from the cosmic casino.

We can estimate the time to remove the asteroid.  Because the motion
is chaotic we can treat the resonance angle as a random variable,
which drives random changes in the eccentricity.  In the case of the
3:1 resonance the changes in $e$ are of order $\delta
e^2 \sim (T_L/T_J) (M_J/M_\odot) e^2$, using the Lyapunov time, $T_L$,
as the interval associated with the random changes ($M_J$, $T_J$, and
$M_\odot$ are Jupiter's mass, Jupiter's orbital period, and the solar
mass, respectively).  The short-term average
eccentricity diffuses to large values in a time
%
%
of about one million years.

In some cases, such as the asteroid Helga in the 12:7 resonance, the
motion is chaotic with a very short Lyapunov time, but the diffusion
time is comparable to or larger than the age of the solar system. When
applied to Helga this theory predicts that the asteroid should survive
for something like 8 billion years. Direct numerical integrations
agree with this prediction\cite{hm}.  We look forward to seeing this
prediction verified observationally.

It is worth noting that the largest gap, that between the 1:1
resonance harboring the Trojan asteroids, and the Hildas in the 3:2
resonance, is the result of the overlap of distinct mean motion
resonances\cite{wisdom80}; for example the 4:3 resonance overlaps with
the 5:4, the 5:4 with the 6:5, and so forth. 

It is believed that most meteorites come from one of two sources, the
3:1 mean motion resonance and the $\nu_6$ secular
resonance\cite{wisdom85a}$^,$\cite{meteorites2}$^,$\cite{mg}. The latter involves a
resonance between the precession frequency of the apsidal line of the
asteroid and the sixth fundamental secular frequency of the solar
system (which is very roughly Saturn's precession frequency).  Roughly
equal numbers of meteorites come from each type of resonance. Until
recently there was one difficulty with this story.  The cosmic ray
exposure ages (a measure of the delivery time) of the stony meteorites
are typically 20 Myr\cite{marti}, where as the delivery time from the
3:1 resonance is much shorter, 1 Myr, as noted above. However, recent
simulations by Vokruhlick\'y and Farinella\cite{vokrouhlicky} suggests
a way out of this dilemma. They suggest that most meteorites, which
are produced by collision between larger bodies in the asteroid belt,
are not injected directly into either the 3:1 or the $\nu_6$
resonance. Rather they are placed in the vicinity of the resonance,
and then dragged  into the unstable region by the Yarkovsky
effect, which arises due to the anisotropic thermal radiation from
those regions of the fragment which are exposed to the sun. With the
addition of this slow dynamical precursor, the story of meteorite
delivery appears to be complete.

{\large {\bf Chaotic Spin-Orbit Resonances}}

One of the most dramatic and important examples of chaos is afforded
by the evolution of planetary spins. This chaos is produced by
resonances between spin and orbit {\em precession} periods. In such
resonances, the asphericity of the planet couples to the
non-axisymmetric perturbation produced by orbital eccentricity or
inclination.  The rather small deviations from perfect sphericity,
typically a few parts in a thousand for larger bodies like planets,
leads to significant exchanges of energy and angular momentum between
orbital and rotational (or spin) motions of satellites and even
planets. Resonances between spin period and orbital period are common
in the solar system, the Moon being a prominent example. The phase
space occupied by such resonances is small, making it unlikely that so
many bodies formed in resonance. This paradox is explained by the fact
that dissipative effects tend to drive bodies into these resonances,
where the motion is stablized\cite{darwin}$^,$\cite{goldreich66}.  As
a result, most resonant satellites are currently deep in their
respective 1/1 spin-orbit resonances, where the motion is regular;
however, in one case, Saturn's satellite Hyperion, the chaotic motion
may have persisted until the present\cite{wpm}.

Because solar tides are so weak, dissipative effects tend to be less
important for planetary spins, allowing for richer dynamics. Seminal
work by Ward\cite{ward73} showed that the angle between the spin and
orbital axes of Mars (the ``obliquity'') varies by a $\pm13.6^\circ$
around its average of $24^\circ$ over millions of years. These
variations are a result of a resonance between the precession of the
spin axes and a combination of orbital precession frequencies;
improvements in orbital models resulted in an increased variation of
$\sim\pm20^\circ$\cite{wr}. A dramatic twist was added by numerical
integrations, which showed that the obliquity of Mars is 
evolving chaotically, and varies over an even larger
range\cite{tw}$^,$\cite{lr}. Such a variation has profound, but as
yet poorly understood implications for climate variation. 
It is likely that the spin axes of Mercury and Venus underwent chaotic
variations in the past\cite{lr}.

The tilt of the Earth, currently 23 degrees, will also increase in the
future, as first pointed out by Ward\cite{ward82}; the Moon evolves
outward under the influence of the tides resulting in a decrease in
the precession rate of the Earth. Eventually the precession rate
becomes resonant with yet another combination of orbital precession
rates.  Once again, numerical integrations show that the Earth's
obliquity will vary chaotically\cite{ljr}$^,$\cite{surgy_laskar}. The tilt
of the Earth's axis may increase to 90 degrees. The effect on our
climate is hard to estimate, but the result is unlikely to be
pleasant.

{\large {\bf Three-Body Resonances}}

Numerical integrations of main belt asteroids show that a substantial fraction
of these bodies have chaotic orbits with rather short Lyapunov times
($\sim {10}^5$ years)\cite{hm}$^,$\cite{MHP98}$^,$\cite{Nesvorney}.
This chaotic motion is not associated with any two-body resonance.
Instead it is the result of the interaction among three bodies: the
asteroid, Jupiter, and either Mars in the inner asteroid belt or
Saturn in the main or outer asteroid
belt\cite{MHP98}$^,$\cite{Nesvorney}.  As with Helga, the Lyapunov
time can be quite short, but the diffusion times are comparable to or
longer than the age of the solar system.  

Three-body resonances arise when one planet, Jupiter say, is perturbed
by a second, Saturn for example. The orbit of Jupiter is then no
longer a simple Keplerian ellipse. The potential experienced by the
asteroid in Jupiter's gravitational field is given by an expression
formally equivalent to the two-body case, but Jupiter's orbital elements
now vary with time. This variation introduces a whole new suite of
frequencies into the potential experienced by the asteroid; in
addition to all the harmonics of Jupiter's period, all the harmonics
of Saturn's orbital period appear as well.  Three-body
resonances have also been considered for the Galilean and Uranian
satellites, as well as in ring systems\cite{tiss}$^-$\cite{aksnes}. 

These new terms in the potential have much smaller amplitudes than two body
resonant terms having the same number of powers of eccentricity.
Three-body resonances are proportional to the product of the masses of the
two perturbing bodies, $M_J$ and $M_S$ in our example, rather than
just $M_J$ in a two body resonance.

The structure of a three-body resonance is similar to that of a two
body resonance---multiple, very narrow components separated in
semimajor axis by an amount that depends on the precession frequencies
of the bodies involved. When the separation is comparable to or
smaller than the width of the individual resonances, the motion in the
immediate vicinity of the resonance will be chaotic. One can then
estimate the Lyapunov time and the diffusion time in the same manner
as for two body resonances.

We have searched the catalogue of asteroids for evidence of gaps at
the location of a number of the stronger three-body resonances. We
have not been able to identify a candidate gap.

We can check the theory in other ways. For example, some asteroids on
the inner edge of the belt (near Mars) could have formed in three-body
resonances involving the asteroid, Mars, and Jupiter. If the diffusion
time of some of these asteroids is comparable to the age of the solar
system, we should be able to find objects that are about to be removed
from the belt due to close encounters, or possibly collisions, with
either Mars or Earth. Morbidelli and co-workers claim that such bodies
have already been discovered, namely the so-called near-Earth
asteroids\cite{morbidelli}. These bodies have been a puzzle for
astronomers, since in their present orbits they have very short
lifetimes, typically millions of years. Since the solar system is
billions of years old, objects with lifetimes of millions of years
should have vanished long ago, unless there is some way to replenish
the supply. The theory of three-body resonances offers a possible
mechanism. In contrast to the situation with Helga, we would prefer
not to test the predictions of the theory, at least in its collisional
aspects.

{\large {\bf Chaos among the giant planets}}

Even trajectories outside but near a resonance can be affected by its
presence.  In the century following Newton's publication of his law of
universal gravitation, astronomers noted that the positions of Jupiter
and Saturn deviated from their predicted positions by some 30 minutes
of arc.  The difference became known as the great inequality.  Laplace
noted that the predictions did not take into account the influence of
Saturn on Jupiter's orbit, and vice versa. On the face of it this neglect
seemed reasonable, because the mass of either planet was less than one
one-thousandth than that of the sun.  Early astronomers had employed
the naive estimate made above, which integrates this tiny force over
the interval between successive conjunctions of the planets, a time
somewhat longer than $T_J$.  But Laplace realized that the orbital
period of Saturn was almost exactly $5/2$ times that of Jupiter
($\frac{2}{T_J}-\frac{5}{T_S} \approx \frac{1}{85T_J}$).  The
perturbations accumulate over a much longer interval ($85 T_J$),
permitting a larger exchange of energy and angular momentum between
the two planets.  The change in the predicted position of the planet on
the sky was roughly $85^2$ times larger than the simple estimate would
indicate. With this result in hand, Laplace was able to reconcile the
observations with the prediction of the law of gravity.

This discovery strongly affected
Laplace's views regarding determinism, reflected in his well known
statement\cite{Laplace}:
\begin{quote}
The present state of the system of nature is evidently a consequence
of what it was in the preceding moment, and if we conceive of an
intelligence that at a given instant comprehends all the relations of
the entities of this universe, it could state the respective
position, motions, and general affects of all these entities at any
time in the past or future.

Physical astronomy, the branch of knowledge that does the greatest
honor to the human mind, gives us an idea, albeit imperfect, of what
such an intelligence would be.
\end{quote}
This view of the world passed into common currency: the clockwork
motion of the planets became the epitome of regularity. 

This view is wrong.  It is wrong because it ignores the delicate
nature of the separatrix, the cradle of chaos.  
In fact the giant planets provide further evidence of the power of three-body
resonances, the subtlety of natural phenomena, and the difficulty of
interpreting the fruits of scientific enquiry. 

The tour-de-force numerical integrations of the outer planets by
Sussman and Wisdom in 1988\cite{SW88} shattered the clockwork.  Using
the ``Digital Orrery'', a parallel computer built specifically for the
the task and now part of the Smithsonian collection, they followed the
motion of the four giant planets and Pluto (as a test particle) over
845~Myr.  To the surprise of all at the time, Pluto's orbit was
chaotic with $T_L\approx10$ million years. 

This revolutionary breakthrough was followed by a re-examination of
the orbital evolution of the full solar system.  Longer integrations
1989 by Laskar of an approximate model of the solar system (the orbits
were averaged, and Pluto was ignored) showed that 
solar system itself was chaotic. Laskar suggested that secular
resonances involving the terrestrial planets were responsible for the
chaotic motion\cite{laskar90}$^,$\cite{laskar92}.  Subsequent integrations
of a complete model by Sussman and Wisdom\cite{sw92} confirmed that
the full solar system was chaotic with $T_L\approx5$~Myr.  Sussman and
Wisdom also found that the four giant planets by themselves appeared
to form a chaotic system.   

The chaos seen amongst the four giant planets arises from a three-body
resonance involving Jupiter, Saturn, and Uranus\cite{mh99}. The orbital
period of Uranus is nearly seven times that of Jupiter. In terms of
the orbital frequency $n=2\pi/T$, the difference $n_J-7n_U$ is equal
to $5n_S-2n_J$. The relevant resonant terms in the potential
experienced by Uranus are proportional to $M_JM_S/(2-5n_S/n_J)$.
Because $(2-5n_S/n_J)\approx1/85$ appears downstairs, it is known as a
``small denominator''; it enhances the potential experienced by Uranus
by a factor of $85$. It was of course exactly this small denominator
that Laplace used to explain the origin of the great inequality, and
which is implicated in the removal of objects from the 2:1 resonance
in the asteroid belt.

The width of a single component of this three-body resonance is tiny,
comparable to the radius of Uranus, but so is the separation between
components; the resonances just overlap. The predicted Lyapunov time
is $T_L\approx10$ million years. Because Uranus is so much less
massive than either Jupiter or Saturn, its orbital angular momentum is
substantially less than that of the two gas giants. Hence it is
subject to the gambler's ruin. The diffusion time (roughly the time
before Uranus is ejected) is $10^{18}$ years, much longer than the
current age of the universe. The theory also predicts the location and
Lyapunov times of other chaotic zones near the present orbit of
Uranus. Detailed numerical integrations verify the presence and
Lyapunov times of all these zones, as well as illustrating the
transitions between libration and rotation in the relevant
resonances. Current computational resources are inadequate to test the
predicted diffusion time.

{\large {\bf Chaos among the terrestrial planets}}

As noted above, numerical integrations of the full solar system, inner
planets included, show evidence of chaos with a Lyapunov time of
$T_L\approx 5$ million years\cite{sw92}$^,$\cite{laskar89}.  Unlike
the outer planets, the source of this chaos has not been convincingly
established.  Laskar pointed to what are called a ``secular
resonances'' between the terrestrial planets as a candidate source of
the chaos\cite{laskar90}.  He found what appears to be an alternation
between circulation and libration in the angles
$\sigma_1 \equiv (\varpi_1-\varpi_5)-(\Omega_1-\Omega_2)$ and
$\sigma_2 \equiv 2(\varpi_4-\varpi_3)-(\Omega_4-\Omega_3)$.  
In these
relations $\varpi_1$ refers to the orientation of Mercury's apsidal
line, which is the line from the sun to the point of Mercury's orbit
closet to the sun, while $\Omega_1$ refers to the orientation of the
Mercury's nodal line; the latter is defined by the intersection of the
orbital plane of Mercury with the orbital plane of the Earth.  Similar
definitions apply to the elements for Venus (2), Earth (3), Mars (4),
and Jupiter (5). 
However, these two resonances do not interact
directly with each other, so by themselves are they unlikely to
produce large scale chaos. Later integrations\cite{laskar92}
identified a third resonance,
$\sigma_3\equiv(\varpi_4-\varpi_3)-(\Omega_4-\Omega_3)$. Laskar found
that it librated while $\sigma_2$ rotated, and vice-versa.  Since
$\sigma_2$ and $\sigma_3$ involve the same degrees of freedom, they
are a more promising candidate for overlapping resonances. Sussman and
Wisdom, employing a more realistic (unaveraged) model, confirmed the
alternate libration and rotation of $\sigma_2$, but never saw
$\sigma_3$ librate; this does not rule out the possibility that the
over lap of the associated resonances causes the chaos. However, their
assessment that ``no dynamical mechanism for the observed chaotic
behavior of the solar system has been clearly demonstrated'' seems
warranted, at least for the terrestrial planets.  Without a clear
identification of the source of the chaos it is not possible to use an
analytic development, such as was used for the outer planets, to
confirm the Lyapunov time and then estimate the time scale for
diffusion of the system.

{\large {\bf Chaos in other planetary systems}}

By analogy with our solar system, multiplanet systems seem likely. In
fact three Jupiter-mass objects orbit Upsilon
Andromeda\cite{Butler}. Any multiple planet system is subject to the
instabilities described in this paper.  All of the known planetary
systems orbiting solar type stars have ages of $10^9$ years or
greater. This fact, together with the Copernican assumption that we
are not observing the system at a privileged time, such as immediately
after a recent planetary ejection, can be used to put limits on the
mass and orbital elements of the planets, quantities which cannot be
tightly constrained or even directly observed using current
techniques\cite{Laughlin}$^,$\cite{Rivera}. Dynamical constraints can
also be placed on the existence of smaller (Earth mass) bodies in
orbits near those of the Jupiter-mass objects. Observational searches
for such low mass companions will have to await improved techniques.

{}

{\bf Acknowledgements} 
We thank Peter Goldreich for helpful conversations.  We also thank
the referees for their assistance in improving the manuscript. This
research was supported by NSERC of Canada and NASA.

\clearpage
\newpage
%
{\large {\bf TEXT OF BOX 1 BEGINS HERE}}

Here we describe a simple model that captures many of the generic
features of conservative or Hamiltonian dynamical systems.  As stated
earlier, at least two oscillators must be interacting or coupled for chaos to
occur.  We can add a second oscillator to the rigid pendulum by
vertically (or horizontally) moving the pivot of the pendulum
periodically.  Much insight can be gained by carefully examining this
and similarly simple models\cite{hm}$^,$\cite{wisdom85a}$^,$\cite{Chirikov}$^,$\cite{LL}

To describe the motion we now need two
angles, say the angle of the pendulum, $\theta$, and the phase of the
pivot, $\psi$.  In addition we need two angle derivatives,
$\dot\theta$ or $p=\dot\theta/\beta$, and $\dot\psi$.  The Hamiltonian is
\begin{eqnarray}
H &=& \frac{1}{2}\beta p^2 + \alpha I + \left[\epsilon + 2 K
\cos\psi \right] \cos\theta\\\nonumber
 &=& \frac{1}{2}\beta p^2 + \alpha I + \epsilon\cos\theta + K \cos(\theta
-\psi) + K\cos(\theta+\psi)\nonumber.
\end{eqnarray}
The first line presents an amplitude-modulated pendulum; the second
presents three phase-modulated pendula. See ref.~\cite{Chirikov} for a
detailed development.  Several constants depend on the physical characteristics of the driven pendulum: $\beta$
is related to moment of inertia of the pendulum; $K$ is a proportional
to the amplitude of the driving; and $\epsilon$ is proportional to the
gravitational acceleration.  The constant $\alpha$ is the driving
frequency or $\dot\psi$.

There are three resonances (cosine terms) which dominate the
motion when their arguments are slowly varying:
$\dot\theta \approx 0$, $\dot\theta \approx \dot\psi$, and $\dot\theta
\approx -\dot\psi$.  We can visualize the trajectories by plotting $p$
versus $\theta$ whenever the driving reaches a specific phase,
$\alpha=0$ for example. The resulting diagram is called a
surface of section.

Figures~7a-7c show surfaces of section for
$\beta=1$, $\epsilon = -1$, and $K = -0.3$.  The panels correspond to
decreasing driving frequencies of $\alpha = 7$, $3$, and $0.3$, respectively.  

Here, the natural oscillation frequency of the pendulum 
is $\sqrt{|\epsilon|\beta} = 1 $.  The resonant islands are located at
$p=0$ and $\beta p=\pm\dot\psi = 
\pm \alpha$.  The center resonant island has a half-width of $\Delta P
= 2\sqrt{|\epsilon|\over|\beta|}=2 $ and the other two have half-widths
of  $\Delta P = 2\sqrt{2 K/\beta}\approx 1.5$. So, for the highest
driving frequency (top panel) the centers of the resonant islands are
separated by more than their widths.  Aside from the addition resonant
islands, two new features appear.  First, two chains of secondary
islands can be seen.  Second, the separatrix of the main island has
broadened into a fuzzy zone.  This zone was traced out by a single
trajectory. Each of the separatrices shows such a chaotic zone,
although they are too small to discern. Nevertheless, what appear to
be invariant curves corresponding to regular trajectories can still be
seen.  That both regular and chaotic trajectories should appear on the same
surface of section is a generic feature of a {\it divided phase
space}\cite{Chirikov60}$^,$\cite{hh}$^,$\cite{Chirikov}.

In the middle panel the driving frequency has been reduced, $\alpha=3$,
such that now the islands, ignoring their interaction, would just
touch each other.  As the driving frequency approaches the natural
oscillation frequency of the pendulum, the small chaotic zone seen in
the top panel envelopes the three resonant islands.  This is the result of the
beginning of {\it resonance overlap}\cite{Chirikov}.

In the bottom panel the driving frequency has been further reduced,
$\alpha=0.3$, such that now the resonances have widths much larger
than their separations.  The motion is chaotic near the separatrix,
which appears as a single thick band surrounding a stable island, and
is surrounded by invariant curves.  As can be seen from the
Hamiltonian, the three resonant terms can also be viewed as a single
resonance with an oscillating width, $\Delta P = 2 \sqrt{|\epsilon + 2
K \cos\psi|\over\beta}$.  The width ranges from $\Delta P \approx 1.3$
to $\Delta P \approx 2.5$, the chosen values of $\epsilon$ and $K$.
Here, the chaotic zone is roughly that region over which the separatrix
sweeps.  This is termed {\it modulational chaos}\cite{tennyson} or
{\it adiabatic chaos}\cite{wisdom85a}.

In discussing the origin of chaotic motion in the main text, we have
treated the region of overlap as though it had no structure.  This
simplification is justified in some cases (for example in high order
resonances in the asteroid belt) and for some applications (for
calculating Lyapunov times), but is not always adequate.  

For example, Wisdom noted that the eccentricity of asteroids in the
3:1 resonance could librate around two different centers, one at the
classical ``forced eccentricity'' $e_f\approx 0.05$ and a second at
$e\approx 0.15$. The chaotic region surrounded the separatrix between
these two points, allowing asteroids to pass from low eccentricities
($\sim0.05$) to high $\sim 0.3$, where they could be removed by
encounters with Mars. Later it was discovered that there was a third
libration center at even higher eccentricity, which was connected to
the first two by chaotic orbits\cite{Ferraz}. This allowed asteroids
to reach $e>0.6$; such objects are removed from the resonance,
typically by plunging into the sun. This motion is made possible by
the resonance overlap, but describing it naturally requires knowledge
of the underlying phase space structure; alternately one can perform
numerical simulations.

Direct brute force numerical
integration indicates that plunging into the sun is the most likely fate of
bodies injected into the 3:1 resonance, with the remainder being
pushed beyond Saturn\cite{Gladman}.  These integrations find a removal
time of about one million years. There is also direct observational
evidence that this interpretation is correct because the boundaries of
the chaotic zone closely match the boundaries to observed distribution
of asteroids\cite{wisdom83}. Similar statements can be made about the
other Kirkwood gaps and gaps in the outer asteroid belt\cite{hm} (see
figure~\ref{tlandte}).

A more realistic model for the 3:1 mean motion resonance, still for a
single perturbing planet, would 
include terms of order $I^2$, and allow for (different) $I$ dependences in the
factors $K$ multiplying the cosine terms involving $\psi$. Such a
model reproduces the two libration centers in the $\psi,I$ motion
found by Wisdom. Because the analytic model that we, and Wisdom,
employ is of low order in $e$, it misses the third libration center
found in reference 25.

Still more complete dynamical models, in particular those that include the
effect of Saturn on Jupiter's orbit, show that secular resonances,
i.e., resonances between the precession periods of the asteroid and
those of Jupiter and Saturn overlap inside low order mean motion
resonance such as the 2:1\cite{hl}$^,$\cite{mm}. This overlap can
enhance the rate of diffusion. Even this is not the entire story;
Jupiter and Saturn are near (but not in) a 5:2 resonance. This ``great
inequality'' enhances the rate of diffusion of asteroids in the 2:1
mean motion resonance with Jupiter\cite{fmr}.

\begin{verbatim}
Figure 7a caption:

Surface of section for the driven pendulum with rapid driving much
faster than the natural pendulum oscillation frequency.  
Top panel of figure in Box 1.

Figure 7b caption:

Surface of section for the driven pendulum with driving frequency
approaching the natural pendulum oscillation frequency. 
Middle panel of figure in Box 1.

Figure 7c caption:
Surface of section for the driven pendulum with driving frequency
less than the natural pendulum oscillation frequency. 
Bottom panel of figure in Box 1.


\end{verbatim}

{\large {\bf TEXT OF BOX 1 ENDS HERE}}
%

\vfil\eject

\begin{figure*}
\caption[penda]{
A rigid pendulum, with angle $\theta$ measured from the vertical. The
pendulum may rotate through $360^\circ$, unlike a simple pendulum
consisting of a mass suspended on a string.
\label{pendulum}}
\end{figure*}

\begin{figure*}
\caption[pendb]{
The motion of a rigid pendulum traces closes curves on phase diagram
showing the angle of pendulum, $\theta$, versus its angular momentum,
$p=ml\dot\theta$. The stable fixed point is at (0, 0).  The separatrix
emanates from the unstable fixed point at  ($\pm\pi$, 0). The shape of
the separatrix resembles a cat's eye.  
\label{pendulum_phase}}
\end{figure*}

\begin{figure*}
\caption[kirkwood]{
The histogram of asteroids as a function of semimajor axis
$a$. Note the distinct deficit of asteroids at, e.g., 2.5, 2.82, and
2.96 AU. Gaps of this type were first noted by Daniel Kirkwood in the
late 1800s. They correspond to orbital resonances with Jupiter; for
example an asteroid placed at 2.5AU will orbit with a period equal to
1/3 of Jupiter's (the 3:1 resonance).
\label{kirkwood}}
\end{figure*}

\begin{figure*}
\caption[Wisdom31]{The orbital eccentricity of an object placed in the
3:1 resonance (at 2.5AU) plotted as a function of time. The initial
eccentricity is small, but chaotic perturbations from Jupiter force
the eccentricity of the asteroid to undergo a random walk, leading to
a net increase in $e$ and the eventual removal of the asteroid from
the solar system.
\label{Wisdom31}}
\end{figure*}

\begin{figure*}
\caption[separation]{The distance between two initially nearly
identical initial conditions for two interacting nonlinear
oscillators. In panel a the motion is near a stable fixed point, while
in panel b the motion starts near an unstable fixed point. In the
latter case (but not in the former) the two initial conditions
separate exponentially with time, and are said to be chaotic.

\label{separation}}
\end{figure*}

\begin{figure*}
\caption[tlandte]{ The location of asteroids in the outer asteroid
belt in the (eccentricity, semimajor axis) plane.  Lyapunov times have
been computed as a function of semimajor axis for several values of
eccentricity.  Points are plotted when the Lyapunov time is below a
threshold value.  These demark the chaotic zones. It is clear that the
known asteroids avoid these regions.
\label{tlandte}}
\end{figure*}

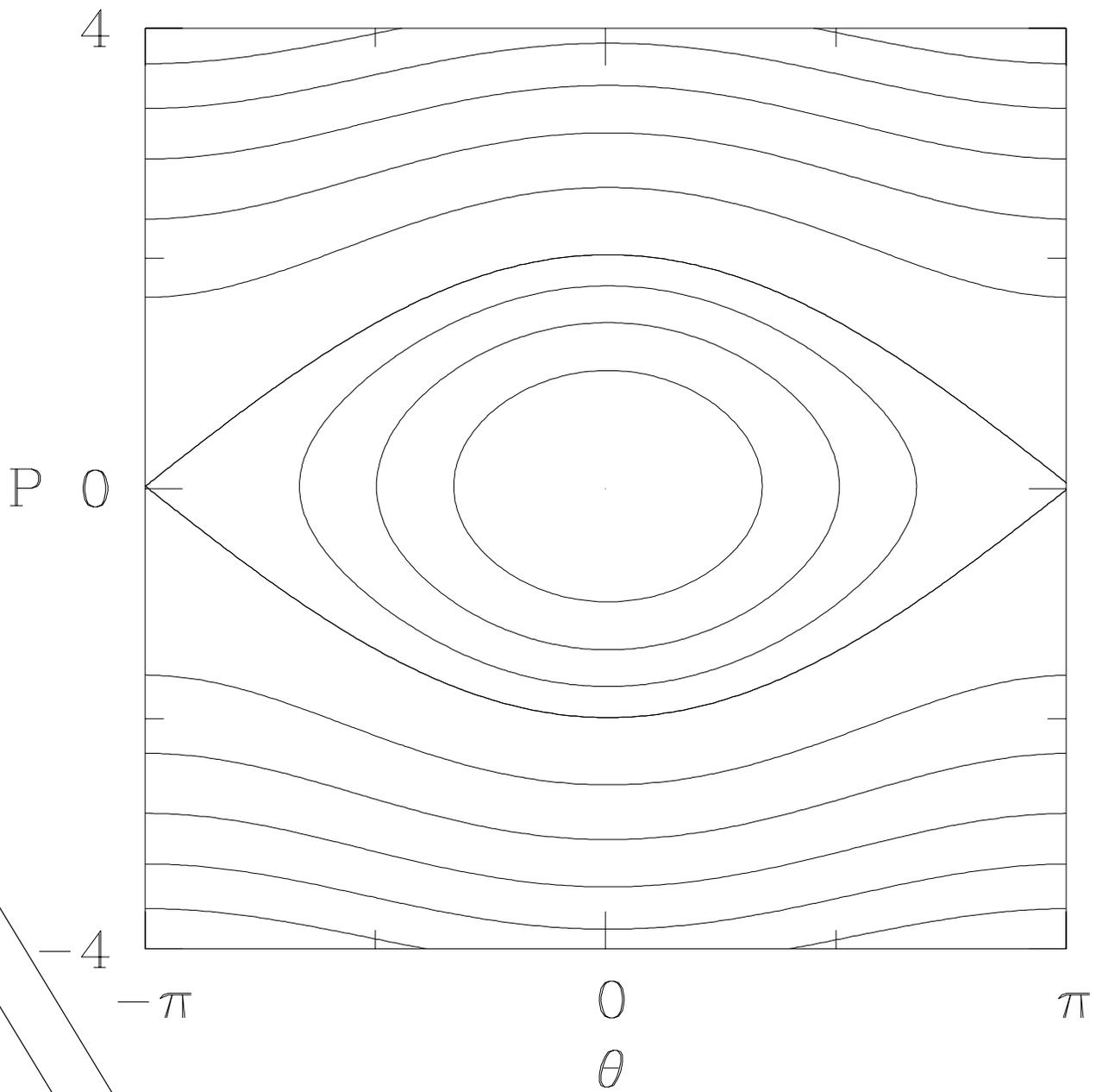
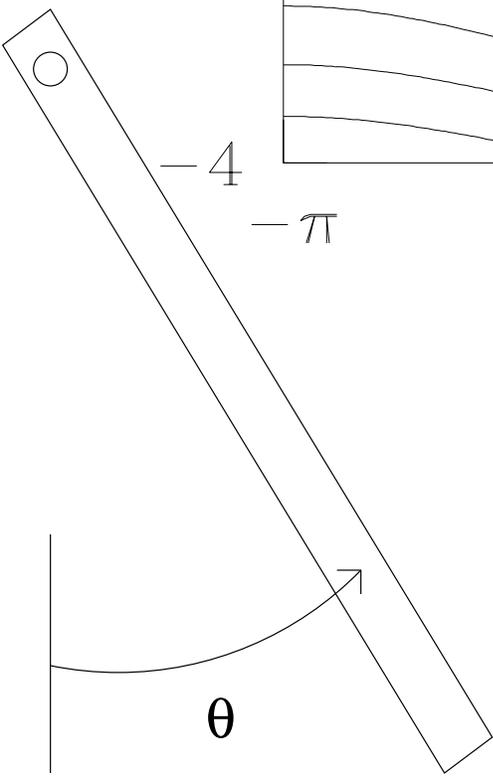
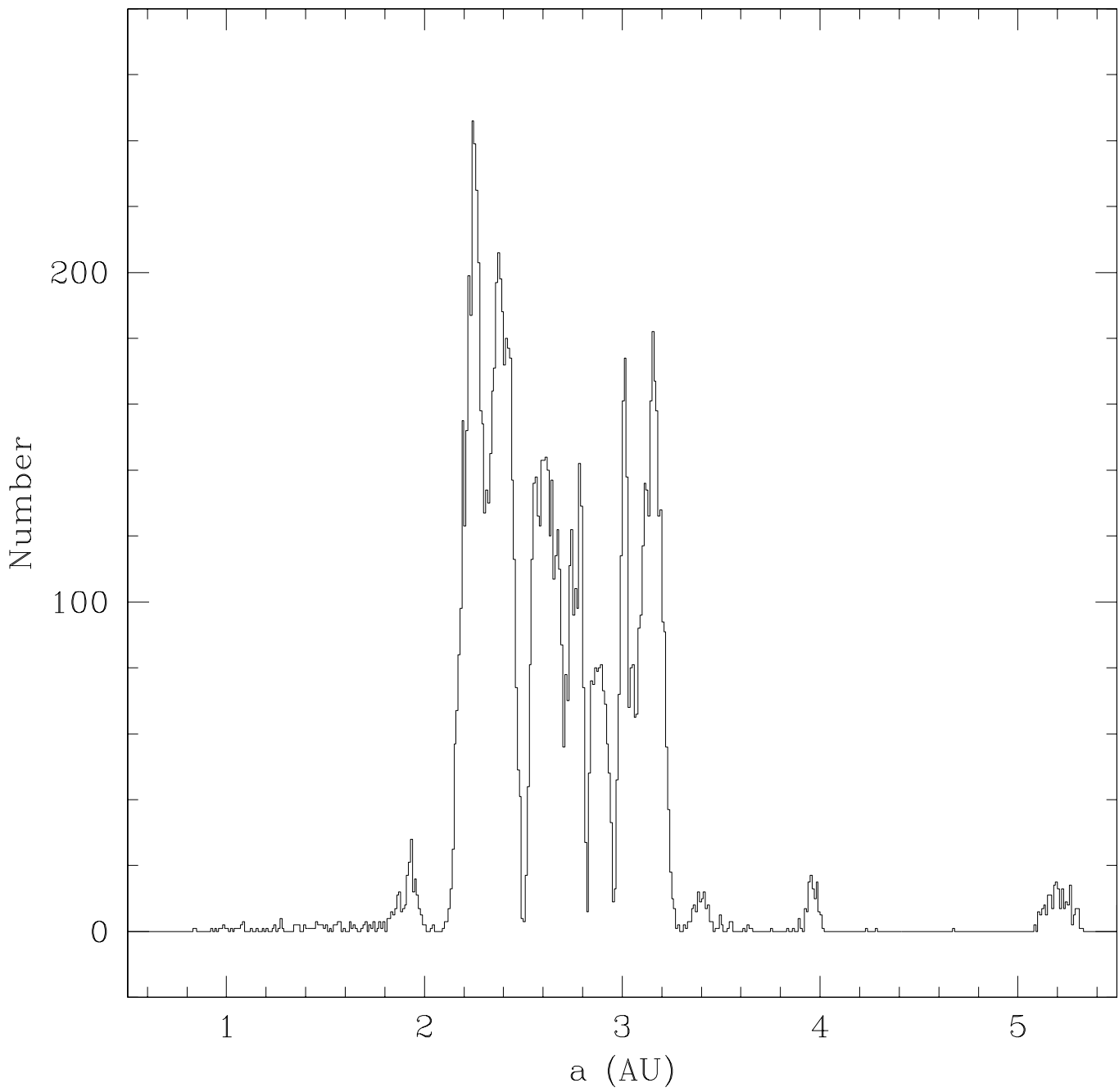
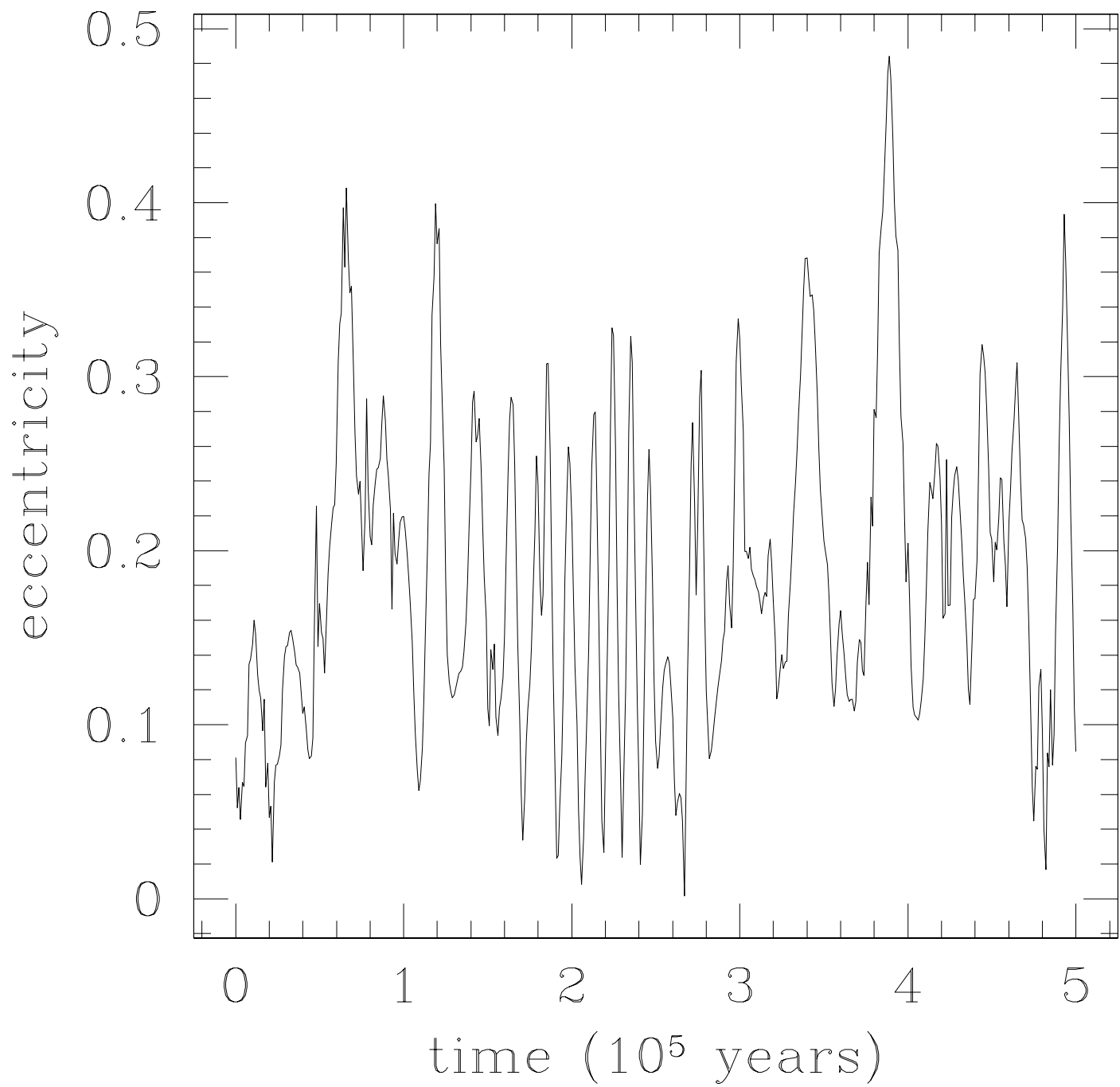
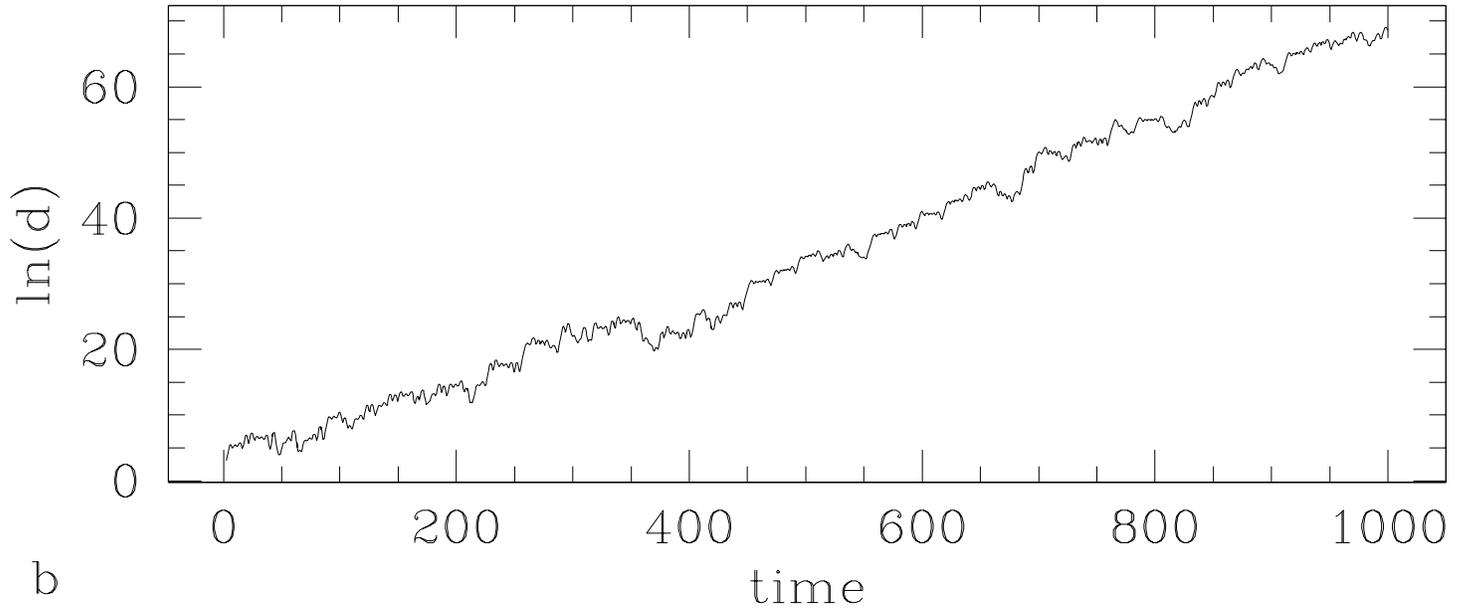
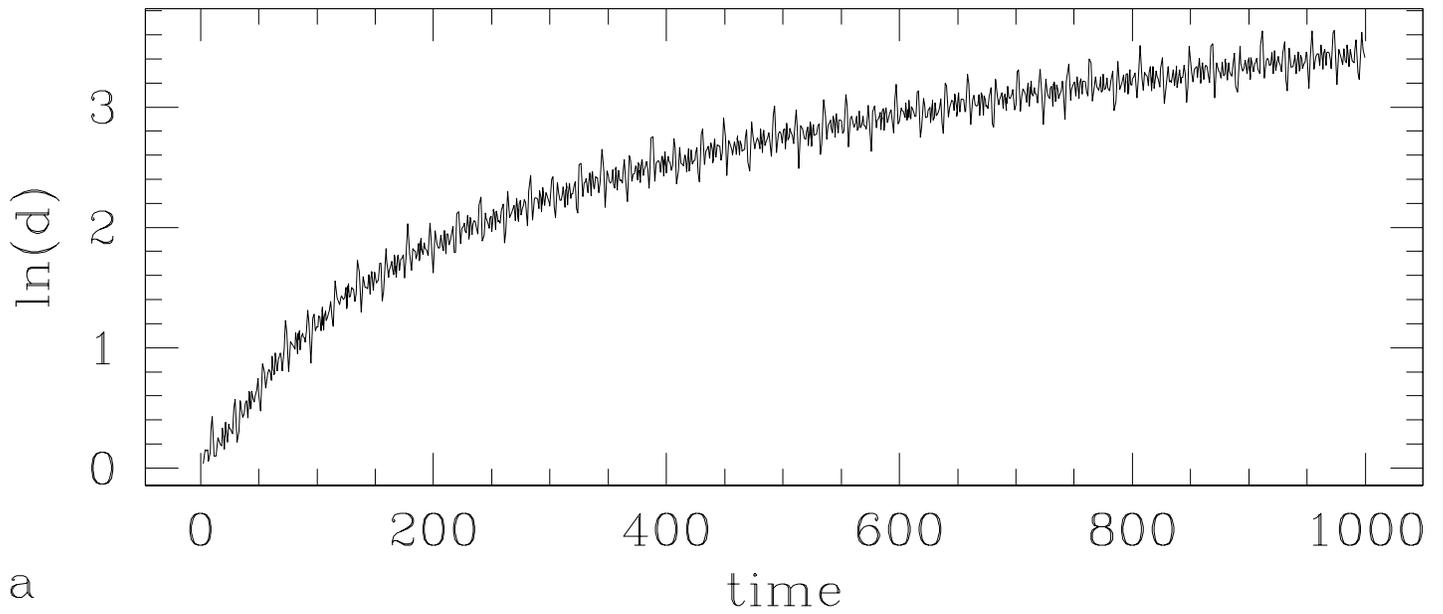
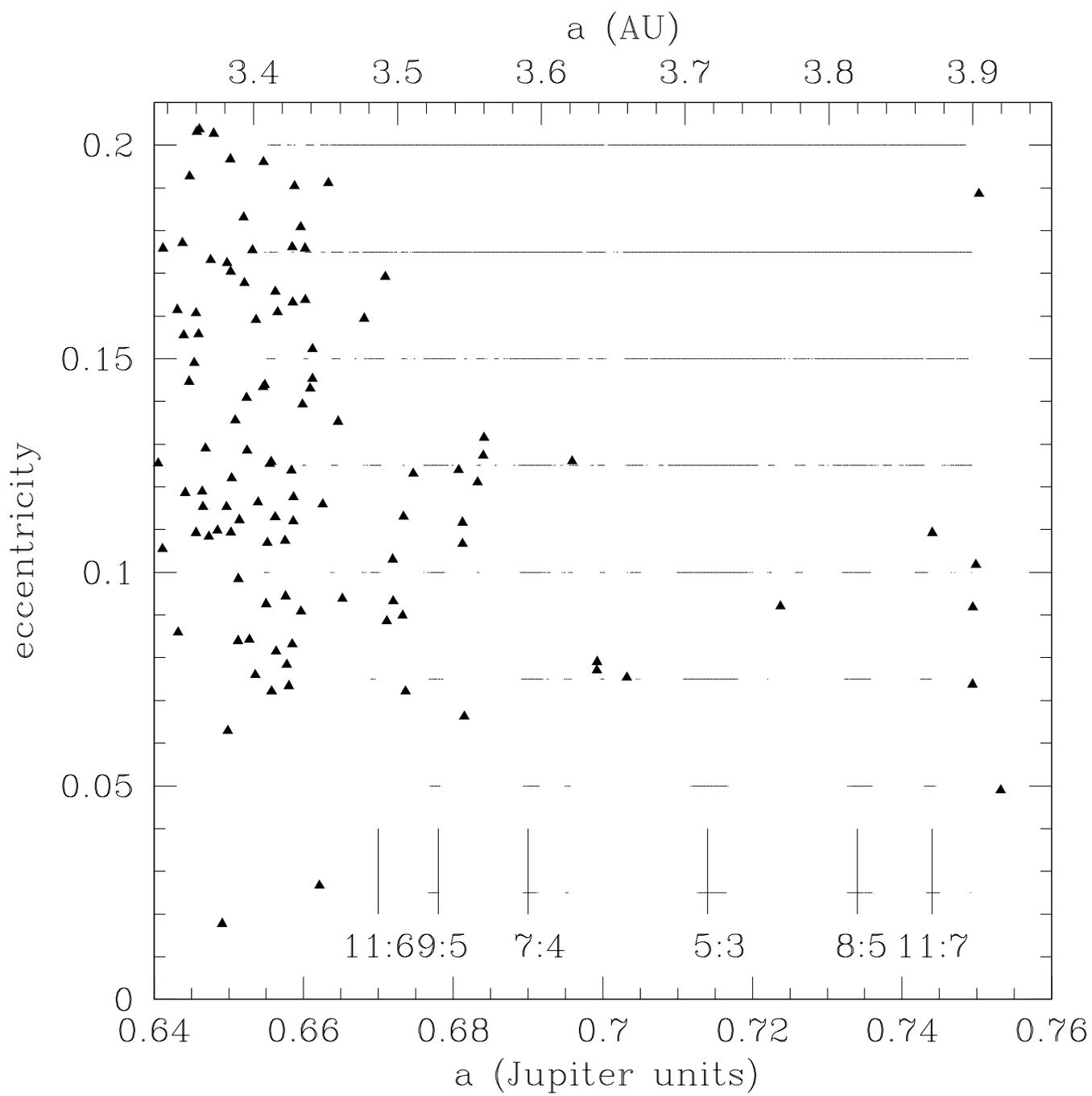
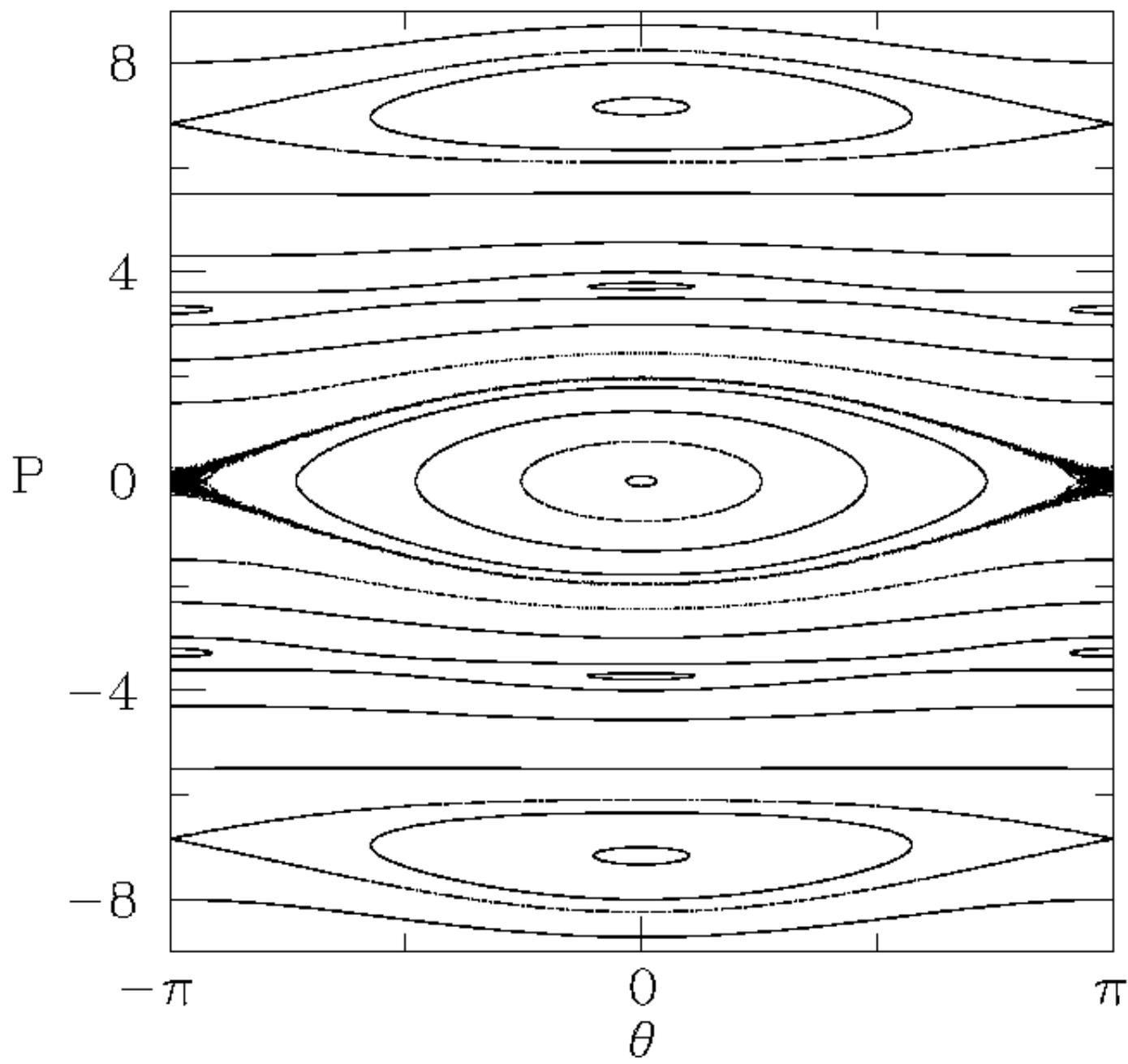
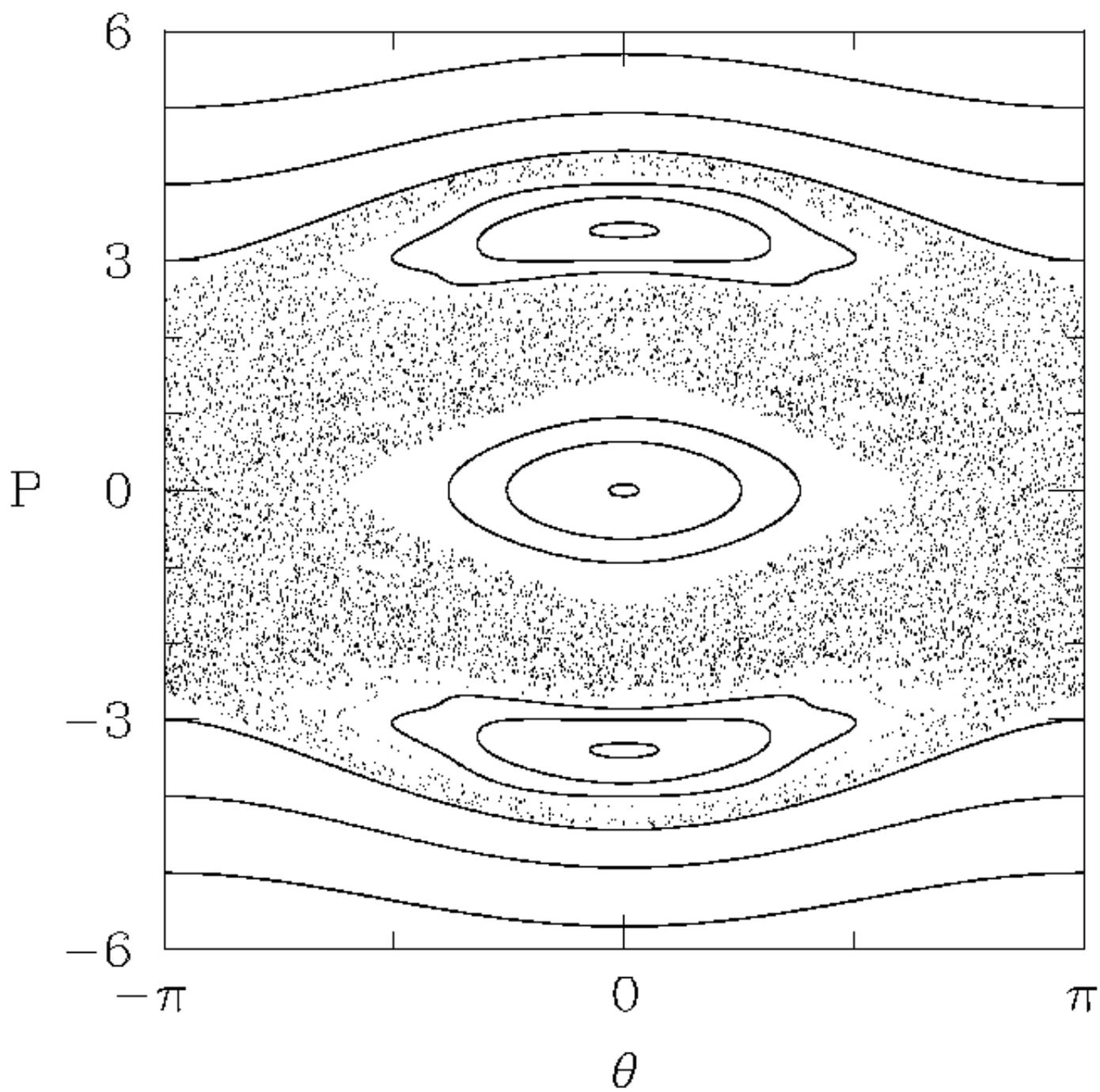
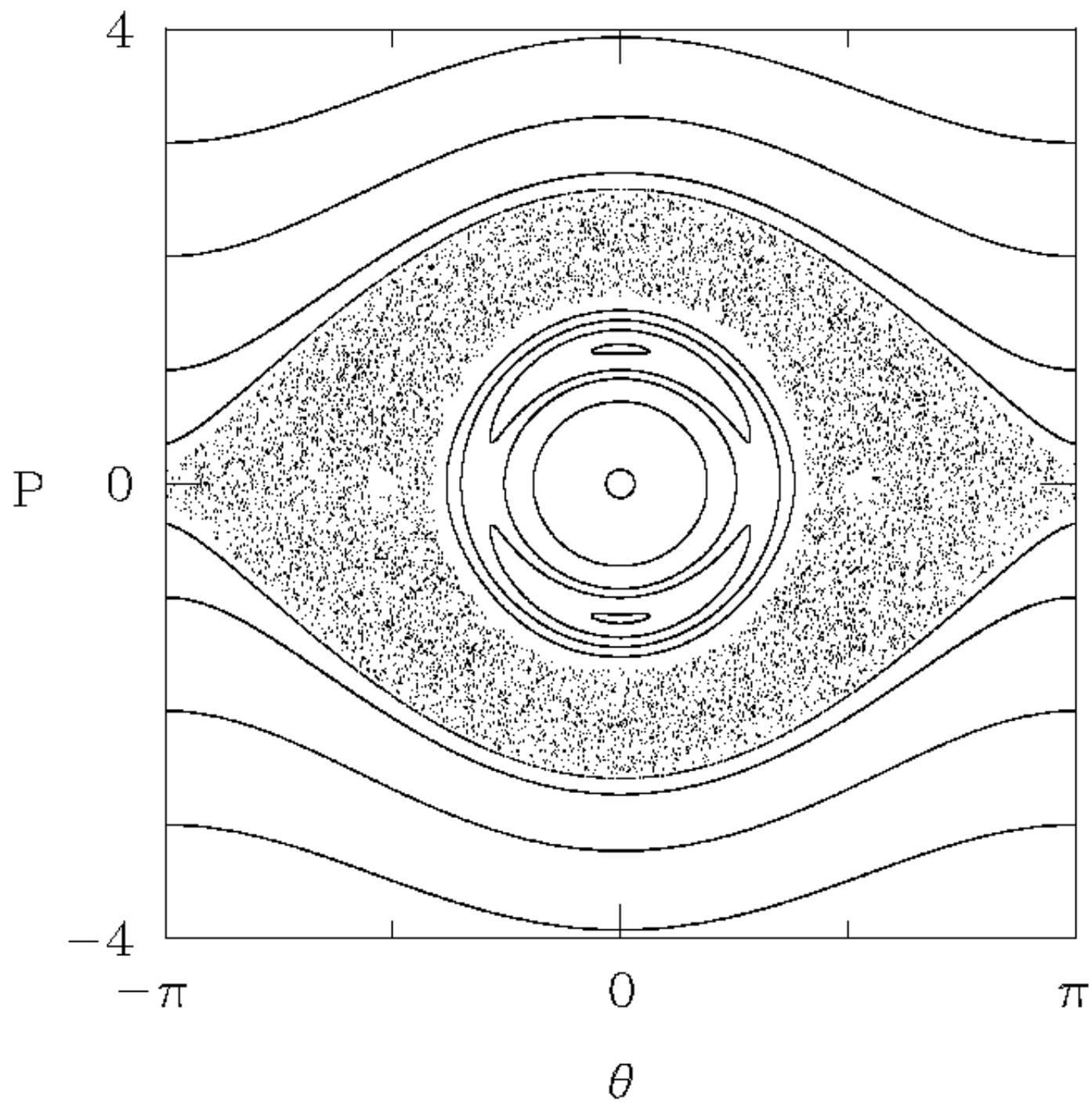
\end{document}